\documentclass[aps,twocolumn,groupedaddress,longbibliography]{revtex4-2}
\usepackage{graphicx}
\usepackage{dcolumn}
\usepackage{bm}
\usepackage{siunitx}
\usepackage{amssymb,amsfonts,amsmath}
\usepackage{xspace}
\usepackage{xcolor}
\usepackage[T1]{fontenc}

\usepackage{hyperref}
\hypersetup{
    colorlinks=true,
    linkcolor=blue,
    filecolor=orange,      
    urlcolor=cyan,
    citecolor = magenta
    }

\newcommand{\SFigPage}{Fig.~S1\xspace}
\newcommand{\SFigqPCR}{Fig.~S2\xspace}
\newcommand{\SFigMeltanAlysis}{Fig.~S3\xspace}
\newcommand{\SFigLengthDependence}{Fig.~S4\xspace}
\newcommand{\SFigMultiplyCenter}{Fig.~S5\xspace}
\newcommand{\SFigMultiplyTerm}{Fig.~S6\xspace}
\newcommand{\STabSequences}{Table~S1\xspace}
\newcommand{\STabPrimers}{Table~S2\xspace}
\newcommand{\STabDNATemplates}{Table~S3\xspace}
\newcommand{\STabDNACombinations}{Table~S4\xspace}
\newcommand{\STabMeltSec}{Table~S5\xspace}
\newcommand{\STabMeltRes}{Table~S6\xspace}
\newcommand{\dna}[1]{\mathrm{#1}}
\newcommand{\cdna}[1]{\mathrm{\overline{#1}}}
\newcommand{\off}{\mathrm{off}}
\newcommand{\D}{\mathrm{d}}
\newcommand{\on}{\mathrm{on}}
\newcommand{\kl}{k_\mathrm{L}}
\newcommand{\R}{\mathrm{R}}
\newcommand{\W}{\mathrm{W}}
\newcommand{\kb}{k_\mathrm{B}}
\newcommand{\Tm}{T_\mathrm{M}}
\newcommand{\mono}{\mathrm{mono}}
\newcommand{\lig}{\mathrm{L}}
\newcommand{\setM}{{\cal M}}
\newcommand{\setE}{{\cal T}}
\newcommand{\setMC}{{\cal M}_\mathrm{cas}}
\newcommand{\setEC}{{\cal T}_\mathrm{cas}}
\newcommand{\error}{\varepsilon}
\newcommand{\errorC}{\varepsilon_\mathrm{cas}}
\newcommand{\Tanneal}{T_\mathrm{anneal}}

\begin{document}
\title{Experimental demonstration of kinetic proofreading\\ inherited in ligation-based information replication}

\author{Hiroyuki Aoyanagi}
\affiliation{Department of Applied Physics, Graduate School of Engineering, Tohoku University, Sendai 980-8579, Japan}
\author{Yasuhiro Magi}
\affiliation{Department of Applied Physics, Graduate School of Engineering, Tohoku University, Sendai 980-8579, Japan}
\author{Shoichi Toyabe}
\email{toyabe@tohoku.ac.jp}
\affiliation{Department of Applied Physics, Graduate School of Engineering, Tohoku University, Sendai 980-8579, Japan}

\date{\today}

\begin{abstract}
We experimentally demonstrate that information replication by templated ligation of DNA strands inherits a kinetic proofreading mechanism and achieves significant error suppression through cascade replication. 
A simple simulation model derived from the experimental results shows that templated ligation has a significant advantage over replication by polymerization for error suppression of long strands.
Specifically, longer chains show lower error rates, significantly distinct from the chain-growth polymerization where errors typically accumulate with chain length.
This mechanism provides a plausible route for high-fidelity replication in prebiotic chemistry and illustrates how physical principles such as nonequilibrium kinetics and network architecture can drive reliable molecular information replication. The approach also offers new strategies for error suppression in biotechnology.
\end{abstract}
\maketitle

\section{Introduction}

Accurate replication of information is essential not only to sustain life but also to advance biotechnology. 
In prebiotic environments, it is not obvious how the fidelity required to maintain genetic information is achieved by primitive error-prone machineries~\cite{Hypercycle, domingo1976, domingo1978, mills1967}. 
In biotechnology, there is a constant demand to reduce errors, including false positives in virus detection by polymerase chain reaction (PCR)~\cite{whiley2005} and ligation-based genotyping~\cite{gibriel2017}.

In general, replication of a nucleic acid sequence is achieved by aligning short complementary strands to a template and joining them.
There are two primary modes of replication, distinguished by how the final product is assembled.
The first is chain-growth polymerization, in which monomers are sequentially added until the final product is formed.
This mode is ubiquitous in both cells and biotechnologies and has been extensively studied in terms of error suppression~\cite{Andrieux2008, sartori2015a, Ouldridge2017,matsubara2023, aoyanagi2023, takahashi2024}.
However, as the sequence length increases, maintaining high fidelity becomes exponentially more challenging, which limits the maximum sequence length that can be reliably replicated, as explained by the theory of error catastrophe~\cite{Hypercycle}.

\begin{figure}[!tb]
\centering{
\includegraphics{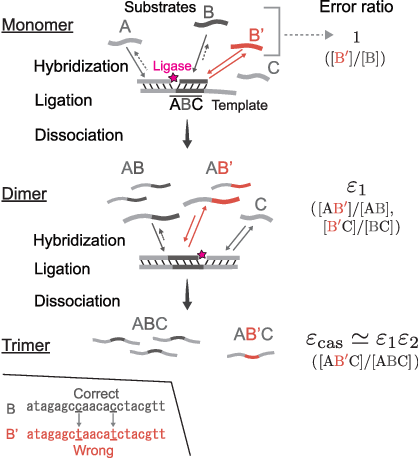}
\caption{Cascade replication by templated ligation. $\dna{A}$, $\dna{B}$, $\dna{B'}$, and $\dna{C}$ are 20-nt DNA oligomers.
Two strands adjacently hybridize (bind) to a template strand and are covalently connected by a ligase enzyme.
Inset: Wrong substrate $\dna{B'}$ has two nucleotide mutations indicated by underlines, which reduce the hybridization stability and also the ligation rate.
 }\label{Fig:1}
}
\end{figure}

An alternative replication mode is templated ligation~\cite{Barany1991, Anderson1983, toyabe2019, kudella2021a, rosenberger2021a, Laurent2024}, where two adjacent short oligomers are joined together on a template~[Fig.~\ref{Fig:1}]. 
Templated ligation has recently attracted growing interest as a key reaction mechanism in prebiotic chemistry, owing to its simplicity and its potential to overcome the error catastrophe via emergent higher-order replication~\cite{Anderson1983, Tkachenko2015, toyabe2019, kudella2021a,rosenberger2021a, matsubara2023, Laurent2024}. 
Templated ligation also has a long history in biotechnology from its use in the ligase chain reaction (LCR) for single-site error detection~\cite{Barany1991,luo1996a} to more recent applications for synthesizing long DNA or RNA strands~\cite{zhou2020a}.
Prominently, information replication by templated ligation can proceed in a cascade manner under nonequilibrium settings of temperature cycling~\cite{toyabe2019, kudella2021a}.
Each ligation step replicates only a partial sequence, but the resulting products can act as substrates in subsequent steps, progressively forming longer sequences.
Over multiple steps, the template sequence is fully replicated in the sequence of the product. 
Recent studies have implicated that this cascade-manner replication enables higher-order replication of information by spontaneously forming cooperative autocatalytic networks~\cite{toyabe2019, Tkachenko2015}. This mechanism implies a frequency-dependent selection, which potentially circumvents the error catastrophe observed in chain-growth polymerization.
However, errors in cascade replication remain largely unexplored.

In this Letter, we demonstrate by experiments and simulations that the cascade replication by templated ligation~[Fig.~\ref{Fig:1}] inherits a kinetic proofreading mechanism for error suppression.
Remarkably, the replication fidelity could increase with sequence length under certain conditions, suggesting the possibility of overcoming the error catastrophe in prebiotic environments and also achieving long sequence replication with high fidelity in biotechnologies.
The kinetic proofreading is a prominent mechanism to achieve fidelity~\cite{hopfield1974,ninio1975, murugan2012}, where cascade reactions make multiple evaluations of errors and thereby increase the fidelity multiplicatively through cascades. 
This mechanism is thought to require a complex molecular mechanism, and its implementation in biotechnology is not straightforward~\cite{mukherjee2024}.
However, our experiments illustrate how physical principles such as nonequilibrium kinetics and network architecture can drive reliable molecular information replication.

\section{Results -- Experiments} 

We prepared a simple model experimental system to measure the replication dynamics and errors in cascaded ligation on a template~[Fig.~\ref{Fig:1}] (see Sec.~\ref{SSec:Experiments} for details). 
We mixed substrate DNA strands, template DNA strands, and thermostable DNA ligase in a total reaction volume of \SI{20}{\micro l} [Sec.~\ref{SSec:DNA}] and repeated a two-step thermal cycle using a thermal cycler [Sec.~\ref{SSec:ligation}].
The cycle consists of a denaturing step at \SI{90}{\degreeCelsius} for \SI{5}{s} and an annealing step at temperatures $\Tanneal$ ranging from 64.8 to \SI{70.5}{\degreeCelsius} for \SI{20}{s}.
The thermal cycling is intended to increase product amounts by suppressing product inhibition.
The substrate strands consist of 20-nt ``monomer'' sequences $\dna{A}$, $\dna{A'}$, $\dna{B}$, $\dna{B'}$, and $\dna{C}$. Wrong substrates $\dna{A'}$ and $\dna{B'}$ contain two-nucleotide mutations relative to $\dna{A}$ and $\dna{B}$, respectively. The 60-nt ``trimer'' template sequence, $\cdna{ABC}$, is complementary to $\dna{ABC}$. 
Here, $\cdna{\,\,\cdot\,\,}$ denotes the complementary sequence and is described from the 3' to 5' ends.
The substrate combinations vary from experiment to experiment.
The initial concentrations of each substrate and template are \SI{200}{nM} and \SI{5}{nM}, respectively. The wrong substrate is mixed so that its fraction becomes 50\%, while maintaining the total concentration at \SI{200}{nM}.
We measured the length-dependent product amounts by denaturing polyacrylamide gel electrophoresis (denaturing PAGE)  [Sec.~\ref{SSec:PAGE}]. Then, we extracted DNA from the gel and quantified the error ratio, defined as the ratio of wrong to correct product amounts, using quantitative PCR (qPCR) [Sec.~\ref{SSec:qPCR}].

\subsection{Replication dynamics}

\begin{figure}[t]
{\centering
\includegraphics{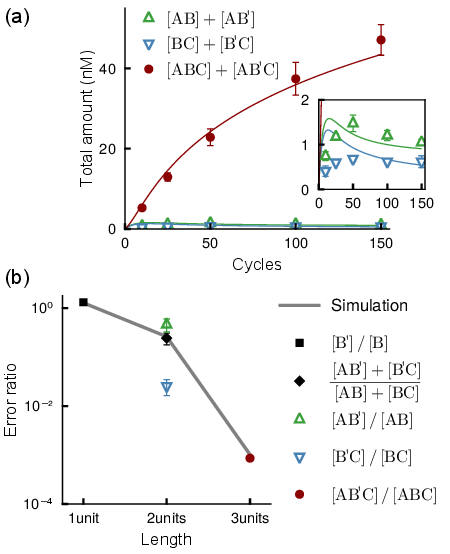}
\caption{Replication dynamics at $\Tanneal=\SI{66.0}{\degreeCelsius}$.
(a) Dynamics of product concentrations: ABC+AB'C (red), AB+AB' (green), and BC+B'C (blue). 
Solid curves are simulations with six fitting parameters, the hybridization rate $k_\on$, the free energy change for the wrong monomer $\Delta G_{\mono,\W}^\circ$, and four types of ligation rates $\kl$ corresponding to the connections AB, BC, $\dna{AB'}$, and $\dna{B'C}$ (see Sec.~\ref{SSec:FittingProcedure}).
The inset is a magnified view of low concentrations. $N=3$ or 4 independent experiments are averaged. 
(b) Single-site error ratios after 150 cycles (symbols) and the simulation (solid curves).
The error bars correspond to the standard errors.
 }
\label{Fig:2}
}
\end{figure}

We measured the replication dynamics with a substrate combination $\setMC = \{\dna{A}, \dna{B}, \dna{B'}, \dna{C}\}$~[Fig.~\ref{Fig:2}].
This is the simplest system to realize the cascade replication of the template sequence information. 
The substrates are ligated on the template and form dimeric intermediate products $\dna{AB}$ ($\dna{AB'}$) and $\dna{BC}$ ($\dna{B'C}$).
These dimers may dissociate from the templates and then rebind to other template strands.
Finally, the dimers serve as the substrates and are ligated to form the final trimeric products $\dna{ABC}$ ($\dna{AB'C}$). 
We observed that dimers accumulate at early cycles and gradually decrease thereafter~[Fig.~\ref{Fig:2}a].
This nonmonotonic dynamics of the intermediate products is a characteristic of cascade replication~[Fig.~\ref{Fig:1}]~\cite{toyabe2019}.
As reactions proceed, monomer concentrations decrease, which results in a decrease in the dimer production rate.
Then, the dimer consumption by trimer formation outpaces the dimer production.

\subsection{Error suppression}

We found a significant decrease in the error ratio with increasing product length~[Fig.~\ref{Fig:2}b]. 
This is in sharp contrast to a \textit{naive} chain-growth polymerization, in which the error ratio at each site is independent of sequence length.

We modeled the experimental system using ordinary differential equations based on mass-action kinetics~[Sec.~\ref{SSec:Simulation}], incorporating hybridization, dissociation, ligation reactions, and thermal cycling with experimentally determined parameters. 
The free energy change $\Delta G^\circ$ associated with the correct hybridization of the sequences involving B was measured by the melting curve analysis.
The ligation rates ($k_{\lig}$), the hybridization rate ($k_\on$), and $\Delta G^\circ$ for wrong hybridization involving $\dna{B'}$ were obtained by fitting the model to experimental data~[Fig.~\ref{Fig:2}a].
The length-dependent dissociation rates ($k_{\off}$) were determined so that they satisfy the detailed balance condition, $k_\on/k_\off=e^{-\Delta G^\circ/\kb T}$, where $\kb$ is the Boltzmann constant. 

The simulation successfully reproduced the experimental results, including the nonmonotonic growth of dimeric strands and decrease in error with the strand length~[Fig.~\ref{Fig:2}, solid curves].
The difference in $\Delta G^\circ$ between monomers B and $\mathrm{B'}$ is approximately $\Delta\Delta G=-1.4\, \kb T$, corresponding to an equilibrium error ratio of $\exp(\Delta\Delta G/\kb T)=0.24$, at \SI{66.0}{\degreeCelsius}~[Sec.~\ref{SSec:FittingProcedure}]. The observed error ratio for the final trimer product, $\mathrm{[AB'C]/[ABC]}=0.00086$, is far smaller than this value, indicating a nonequilibrium discrimination mechanism.
We use a square bracket $[\,\cdot\,]$ for the molar concentration throughout the paper.
Error ratio of monomer is larger than one because of the greater consumption of \(\dna{B}\) over \(\dna{B'}\). The difference in the error ratios of AB and BC mostly originates from the difference in the ligation rates. Mutations in the strand with its 3'-end ligated greatly reduce the ligation efficiency compared to the mutations in the other strand~\cite{luo1996a}. 

\subsection{Error suppression at individual steps}
\begin{figure}[t]
{\centering
\includegraphics{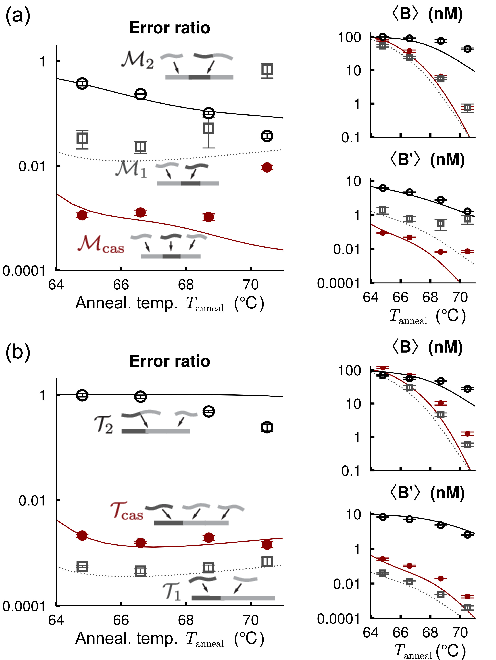}
\caption{Single-site error ratios (left) and product concentrations (right) after \SI{150}{cycles} for the middle (a) and terminal (b) errors.
$\langle \mathrm{B} \rangle$ and $\langle \mathrm{B'} \rangle$ denote the amount of the products containing B and $\dna{B'}$ units, respectively (e.g. $\langle\mathrm{B}\rangle=[\dna{AB}]$ for $\setM_1$ and $\langle\mathrm{B}\rangle=[\dna{ABC}]$ for $\setM_2$ and $\setMC$). Error bars indicate standard errors. Curves in (a) are simulation results. The parameters are obtained by fitting the dynamics (see Fig.~\ref{Fig:2}a).
For (b), $\kl$ for the $\dna{A'B}$ connection was additionally estimated. We assumed that $\kl$ and $k_\on$ do not depend on $\Tanneal$.
}
\label{Fig:3}
}
\end{figure}

To investigate how the error propagates through cascade steps, we measured error incorporation at individual steps using substrate combinations $\setM_1 = \{ \dna{A}, \dna{B}, \dna{B'} \}$ and $\setM_2 = \{ \dna{A}, \dna{BC}, \dna{B'C} \}$, corresponding to the first and second steps of the cascade, respectively, in addition to $\setMC$~[Fig.~\ref{Fig:3}a].
We denote the corresponding single-site error ratios of the final products ($[\dna{AB'}]/[\dna{AB}]$ for $\setM_1$ and $[\dna{AB'C}]/[\dna{ABC}]$ for $\setM_2$ and $\setMC$) as $\error_1$, $\error_2$, and $\errorC$, respectively.
We found that the error ratio with cascade reaction, $\errorC$, is significantly lower than those of its elementary steps, $\error_1$ and $\error_2$.
Importantly, $\errorC\ll\error_2$ despite the fact that the cascade replication with $\setMC$ and the one-step replication with $\setM_2$ produce the same final products $\dna{ABC}$ and $\dna{AB'C}$.
This implies that cascade replication implements an error suppression mechanism.
With cascade replication, dimer products dissociate after ligation in the first step.
The amounts of these dimers are strongly biased toward AB and BC over $\dna{AB'}$ and $\dna{B'C}$, since $\error_1 \ll 1$.
Because these dimers serve as substrates in the second step, the bias propagates to the next step, further reducing the error ratio in a multiplicative manner.

We compared the product $\error_1 \error_2$ to the observed cascade error ratio, $\error_\mathrm{cas}$~[{\SFigMultiplyCenter}a~\cite{Note1}].
Although there is a slight deviation in absolute values, the similarity between the two quantities suggests that $\error_\mathrm{cas}$ can be reasonably approximated by $\error_1 \error_2$.
This trend indicates that the concentration bias established in the first replication stage is effectively propagated to the second stage.

We found that $\error_1$ has a minimum around \SI{66.6}{\degreeCelsius} and $\error_2$ has a minimum at $\Tanneal$ larger than \SI{70.5}{\degreeCelsius}.
This dependence occurs because the difference in hybridization stability between correct and wrong sequences is most pronounced at $\Tanneal$ close to their melting temperatures: $T_\mathrm{m} = \SI{64.5}{\degreeCelsius}$ for B and \SI{75.6}{\degreeCelsius} for BC~[Sec. \ref{SSec:Melting}].
These trends are quantitatively reproduced by simulations~[curves in Fig.~\ref{Fig:3}a]. Here, we assumed the ligation rate $\kl$ and hybridization rate $k_\on$ are constant, independent of $\Tanneal$ for simplicity.
The deviation between simulations and experiments is significant at high $\Tanneal$, possibly due to the above approximation of the temperature dependence of $\kl$ and $k_\on$, and implies the applicability limit of the simulation model.

\subsection{Kinetic proofreading}

We derive the factor that determines the efficacy of error discrimination.
The errors occur when the wrong substrates are fixed by ligation before dissociating from the templates.
The dissociation occurs spontaneously in the reaction step with rates $k_\off$ or is forced in the denaturing step after the reaction step, with a duration of $\tau=\SI{20}{s}$.
Assuming that binding takes place immediately after the reaction step starts, the average duration until the dissociation becomes $(1-e^{-k_\off\tau})/k_\off$.
We define an effective dissociation rate as its inverse $\widetilde k_\off\equiv k_\off/(1-e^{-k_\off\tau})$.
Hence, error discrimination is expected when a criterion, ${k_\lig}/\widetilde k_\off>{k_\lig'}/\widetilde k_\off'$, is satisfied.
Here, $k_\lig'$, $k_\off'$, and $\widetilde k_\off'$ are the rates for the reactions involving wrong substrates.
We define an error discrimination factor $f$ and rewrite this criterion as
\begin{equation}
    f \equiv \frac{k_\lig}{k_\lig'}\frac{\widetilde k_\off'}{\widetilde k_\off}>1.
    \label{Eq.condition}
\end{equation}
The fitting of simulation curves to the experiments gives $k_\lig/k_\lig'\simeq 11.7$ for the ligation between A and $\dna{B\,(B')}$ and $113$ for that between $\dna{B\,(B')}$ and C, independent of the substrate length~[Sec.~\ref{SSec:FittingProcedure}].
The experiments also give $\widetilde k_\off'/\widetilde k_\off\sim 7$ for monomers and $\sim 1$ for dimers~[Sec.~\ref{SSec:TempDepMid}].
Since most dimers do not dissociate during the reaction step because of their stable hybridizations, $\widetilde k_\off'$ and $\widetilde k_\off$ of dimers are similar to $1/\tau$, giving a ratio close to 1. 
Thus, we expect significant error discrimination with $\setM_1$ and $\setM_2$ since both exhibit $f\gg 1$.
These arguments indicate that error is kinetically discriminated through the slower ligation and faster dissociation of wrong substrates.
The error ratio after each cascade step is determined by the product of $f$ and the concentration ratio of the correct to the wrong substrates.
As this happens at each cascade step, the concentration becomes increasingly biased towards the correct one in a multiplicative way through the cascades.

\begin{figure}[t]
{\centering
\includegraphics{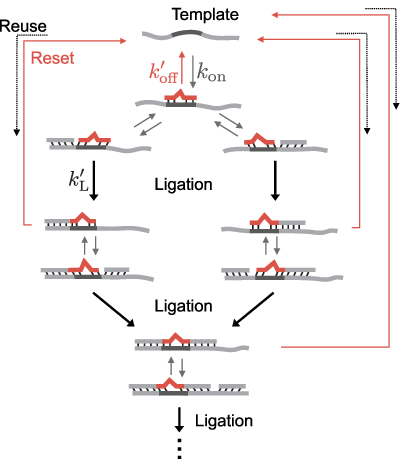}
\caption{Kinetic proofreading in ligation. The diagram is drawn with a focus on a specific template strand, and only the main reaction pathway involving a wrong substrate is shown. The loop structures that reset the state are characteristic of the kinetic proofreading~\cite{murugan2012}. The ``Reuse'' process indicates the hybridization of intermediate products.
}
\label{Fig:4}
}
\end{figure}

This multiplicative error suppression mechanism with kinetic error discrimination is called the kinetic proofreading mechanism~[Fig.~\ref{Fig:4}]~\cite{hopfield1974, ninio1975, murugan2012}.
In our system, the virtually irreversible process of ligation provides the free-energy dissipation necessary for the discrimination. From each intermediate state, the state is reset by the dissociation of intermediate prodcuts. 
The dissociated intermediates may hybridize to the template again.
This reusing, which is not implemented in the original Hopfield-Ninio kinetic proofreading~\cite{hopfield1974, ninio1975}, may increase the replication speed~\cite{murugan2012}.

\subsection{Terminal error}

We evaluated also the terminal error by introducing $\dna{A'}$ instead of $\dna{B'}$; $\setEC = \{\dna{A}, \dna{A'}, \dna{B}, \dna{C}\}$, $\setE_1 = \{\dna{A}, \dna{A'}, \dna{B}\}$, and $\setE_2 = \{\dna{AB}, \dna{A'B}, \dna{C}\}$~[Fig.~\ref{Fig:3}b].
The cascade replication with $\setEC$ shows significantly smaller errors than the one-step replication with $\setE_2$, implying the effectiveness of the cascade replication in reducing terminal errors.

We found a correlation between the product $\error_1 \error_2$ and $\error_\mathrm{cas}$~[{\SFigMultiplyTerm}~\cite{Note1}], supporting the cascade replication.
Interestingly, $\error_2 \simeq 1$, particularly at low $\Tanneal$, and consequently, $\errorC\simeq \error_1$. 
This is because $\widetilde k_\off' / \widetilde k _\off \sim 1$, due to the stable hybridization of dimers~[Sec.~\ref{SSec:TempDepMid}], and $k_\lig' / k_\lig \sim 1$, as the mutation is located far from the ligation site and thus does not significantly inhibit the reaction.
Note that there are two reaction pathways for $\setEC$: $\dna{A' + B + C} \to \dna{A'B + C} \to \dna{A'BC}$ and $\dna{A' + B + C} \to \dna{A' + BC} \to \dna{A'BC}$. In the latter, error discrimination does not occur at the first ligation step, potentially compromising the overall stringency of error suppression in cascade replication as $\errorC>\error_1$.
Indeed, the simulation shows that about 60\% of reactions follow this less discriminative pathway under the present conditions.

\section{Results -- Numerical simulations}
We investigate general situations with template strands of length $l \geq 4$, which may involve a cascade of more than two steps, by simulation [Sec.~\ref{SSec:SingleSite}]. We built the simulation model by extending the rate equations used for the fitting in Fig.~\ref{Fig:2}, but with a different parameter set to evaluate the essential feature of the error suppression mechanism in long sequences. 
For simplicity, thermal cycling was not introduced.
For comparison, we also modeled a chain-growth polymerization, where the hybridization and dissociation of only monomers are considered.
This was implemented by a constraint that, once an intermediate is formed, it remains bound to the template without dissociation.

\subsection{Single-site error}

\begin{figure}[t]
{\centering
\includegraphics{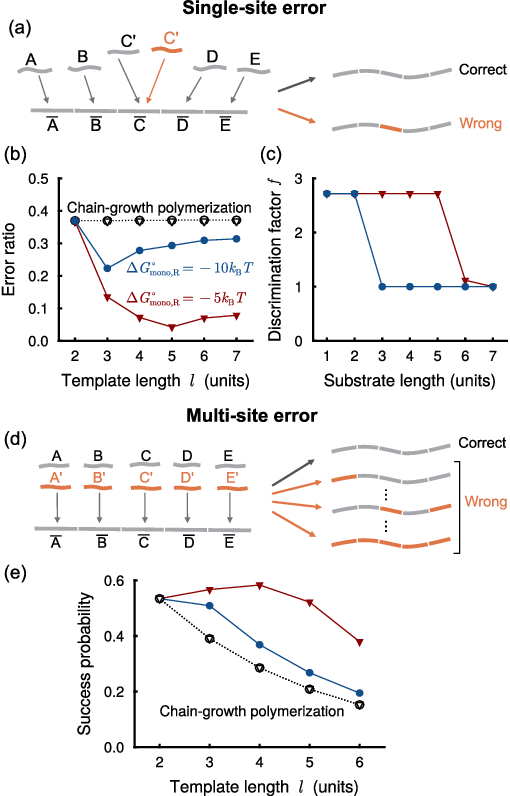}
\caption{Simulations of long-template replications with single-site errors (a--c) and multi-site errors (d, e).
Cascade replication with stable monomer hybridization (blue closed circle) and less stable one (red closed triangle) with $\Delta \Delta G = 1\, \kb T$, and corresponding conditions of chain-growth polymerization (open circle and triangle) are compared in (b), (c), and (e)~[Sec.~\ref{SSec:SingleSite}]. (a) Schematic of single-site error replication, where an error may occur at the central position in a template of length $l$. The example of $l=5$ is illustrated. 
(b) Error ratio. 
 The error ratio for chain-growth polymerization is given by the product of $\exp(-\Delta\Delta G / \kb T)$ and $k_{\mathrm{L}}' / k_{\mathrm{L}}$ and is 0.37, where $\Delta\Delta G=1\,\kb T$ and $k'_{\mathrm{L}} / k_{\mathrm{L}}=1$ are assumed.
(c) Discrimination factor $f$ defined by Eq.~\eqref{Eq.condition} for each length of substrates.
(d) Schematic of multi-site error replication. Errors can occur at any site.
(e) Success probability of replication in the multi-site error case.
}
\label{Fig:5}
}
\end{figure}

First, for simplicity, we simulate a simple case where an error occurs only at the central site of the sequence (i.e., the $(l/2)$-th site for even $l$)~[Fig.~\ref{Fig:5}a].
Two hypothetical parameter sets were used: one with high monomer hybridization stability ($\Delta G_\mathrm{mono, R}^\circ=-10\,\kb T$) and another with low stability ($-5\,\kb T$), while keeping the energy difference between correct and wrong monomers fixed at $\Delta \Delta G = 1 \kb T$, which defines the ratio $k'_\off/k_\off=\exp(\Delta\Delta G/\kb T)=2.7$.
In experiments, the monomer hybridization stability can be controlled by changing, for example, the monomer length, sequence, and temperature.
Here, thermal cycling was not implemented for simplicity, hence, we used the total reaction time as $\tau$ to calculate the effective dissociation rates $\widetilde k_\off$ and $\widetilde k_\off'$.

Across all tested conditions, cascade replication consistently exhibits lower error ratios than chain-growth polymerization for $l \geq 3$~[Fig.~\ref{Fig:5}b].
In chain-growth polymerization, the error ratio is given by the discrimination factor $f$ in Eq.~\eqref{Eq.condition} with the parameters for monomers, since dissociation of intermediates is not considered, and the fraction of the wrong monomers is set to be 50\%.
In cascade replication by ligation, the error ratio steeply decreases between $l=2$ and 3 and then slowly increases with high monomer stability or further decreases with low stability.
Thus, error suppression by cascade replication remains effective for longer sequences, particularly under conditions where hybridization is less stable.

The dependence of the error ratio on $l$ is explained by the discrimination factor $f$~[Fig.~\ref{Fig:5}c].
Short strands have large $k_\off$ and hence $\tilde k_\off\simeq k_\off$, leading to $f\simeq k'_\off/k_\off= 2.7$.
Long strands have small $k_\off$ and, therefore, $\tilde k_\off\simeq \tau$, leading to $f\simeq 1$.
The threshold length dividing the two regimes depends on $\Delta G_\mathrm{mono, R}^\circ$.
As replication progresses, intermediate products become longer and more stable, and hence the discrimination becomes less effective.
Another important factor to consider is that longer templates are replicated via multiple pathways (e.g., A + BCD, AB + CD, or ABC + D for replicating ABCD). The efficacy of the discrimination depends on the pathway, and the number of possible pathways rapidly increases with the length. Therefore, characterising error suppression in the replication of long strands is not straightforward.

\subsection{Multi-site errors}

We further examined a more generalized case in which mutations can occur at any position of the product sequence~[Fig.~\ref{Fig:5}d], using the same hybridization stability conditions tested in Fig.~\ref{Fig:5}b.
In this scenario, replication of a template of length $l$ generates up to $2^l$ distinct product species~[Sec.~\ref{SSec:SingleSite}].
We define the success probability as the fraction of the correct (mutation-free) product among all $2^l$ final products~[Fig.~\ref{Fig:5}e].
We found that the cascade replication achieves higher success probabilities than chain-growth polymerization for $l \geq 3$.
In both, the success probability decreases with $l$, consistent with the trends observed for the single-site error~[Fig.~\ref{Fig:5}b].

\section{Discussion}

Templated ligation is attracting interest as a plausible primitive replication mechanism, capable of generating rich replication dynamics despite its simplicity~\cite{Anderson1983, Tkachenko2015,toyabe2019, kudella2021a, rosenberger2021a, matsubara2023, Laurent2024}. Previous studies have suggested that ligation is effective in preserving sequence information through higher-order replication in situations where both sides of the double strands are replicated~\cite{toyabe2019, Tkachenko2015, kudella2021a}.
In our experiments, we focused on the elementary replication steps, where only one strand of the double-stranded DNA is replicated.
We demonstrated that templated ligation inherits a kinetic proofreading mechanism that suppresses errors through a cascade of kinetic error discrimination steps [Fig.~\ref{Fig:4}] and leads to reduced errors with length in certain length ranges [Fig.~\ref{Fig:5}e].
This has prominent implications for the study of the origin of life, where replication of a long sequence with fidelity is necessary for developing complex functions.
Importantly, this mechanism does not require any external intervention, such as dilution or thermal cycling, as shown in Fig.~\ref{Fig:5}.
In the present experiments, we introduced thermal cycling to prevent the product inhibition, in which ligated products occupy the template and prevent it from templating further ligations.
Thermal cycling is effective in increasing the product amounts but reduces the error discrimination effect by introducing $\tau$, which reduces $f$ given by Eq.~\eqref{Eq.condition}, implying a trade-off between amount and fidelity.

To characterize the essence of error discrimination, we used relatively long-sequence monomers with symmetric mutations in the middle of the sequence, which minimizes the influence of the ligase-specific bias.
Mutations near either monomer-strand end will slow down the ligation at this end due to reduced efficiency in forming the phosphodiester bond~\cite{pritchard1997} or due to an end flap~\cite{takahashi2024}. 
The effect of such asymmetric ligation speed as well as non-enzymatic ligations~\cite{zhou2020a} on the kinetic proofreading mechanism remains unclear and is a topic for future study.

From a bioengineering perspective, the kinetic proofreading by the cascade ligation has potential practical applications, including the genotyping by the so-called ligation detection reaction~\cite{Barany1991} and the production of long strands with fidelity for DNA storage technologies~\cite{doricchi2022}.
In addition, the concept of hierarchical assembly, which resembles cascade replication, appears in the field of macromolecular self-assembly~\cite{wei2024}. Our findings may offer a new perspective on these systems by providing a methodology to build more accurate assemblies.

\section{Methods}
\label{SSec:Experiments}


\subsection{DNA sequences} 
\label{SSec:DNA}
DNA oligonucleotides with either a reverse-phase cartridge or polyacrylamide gel electrophoresis (PAGE) purification grade were purchased from IDT or Eurofins Genomics (see \STabSequences for the sequences~\footnote{See Supplemental Material at [URL will be inserted by publisher] for supplementary methods and results}). Substrates B, $\dna{B'}$, C-X, BC-X, and $\dna{B'C}$-Y were phosphorylated.
The template $\cdna{ABC}$ (written from 3' to 5' ends) consists of 74 bases, consisting of three 20-base domains for substrate hybridization and 7-base overhangs at both ends. These 7-base overhangs are essential for distinguishing the template DNA from the products in PAGE analysis.  
The substrates were U-A, B, $\dna{B'}$, and C-X for $\setMC$, U-A, B, and $\dna{B'}$ for $\setM_1$, and U-A, BC-X, $\dna{B'C}$-Y, for $\setM_2$.
For the terminal error case, we used U-A, $\dna{U'}$-$\dna{A'}$, B, and C-X for $\setEC$, U-A, $\dna{U'}$-$\dna{A'}$, and B for $\setE_1$, and U-AB, $\dna{U'}$-$\dna{A'B}$, and C-X for $\setE_2$.
The sequence $\dna{A'}$ and $\dna{B'}$ contains a two-base mutation relative to substrate A and B, respectively.
Tag sequences U, $\dna{U'}$, X, and Y correspond to primer-binding regions used in the quantitative PCR (qPCR) experiments [Sec.~\ref{SSec:qPCR}].

\subsection{Ligation experiment}
\label{SSec:ligation}
The reaction mixture had a total volume of \SI{20}{\micro \liter}, containing \SI{640}{units\per\milli\liter} of thermostable Taq DNA ligase (New England Biolabs, MA, USA), 1× Taq DNA Ligase Buffer, 5 \si{nM} of the template $\cdna{ABC}$, and each substrates at the total substrate concentration of \SI{200}{\nano M}. When A and $\dna{A'}$ or B and $\dna{B'}$ are mixed, each substrate was added at a concentration of \SI{100}{\nano M} so that $[\dna{A}]+[\dna{A'}]=\SI{200}{nM}$ or $[\dna{B}]+[\dna{B'}]=\SI{200}{nM}$.
The thermal cycling was done with an initial \SI{5}{s} denaturation step at \SI{90}{\degreeCelsius}, followed by \SI{20}{s} incubation steps at five different temperatures: 64.8, 66.0, 66.6, 68.7, and \SI{70.5}{\degreeCelsius} under a real-time PCR cycler (CFX96, Bio-Rad).
After the thermal cycling, an equal volume of loading buffer, consisting of \SI{10}{M} urea and \SI{240}{\milli M} ethylenediaminetetraacetic acid (EDTA) dissolved in 1$\times$ Tris-borate-EDTA buffer (TBE buffer, T3913, Sigma-Aldrich), was added to ensure the complete stop of the reaction.

\subsection{Measurement of product amount}
\label{SSec:PAGE}
Following the thermal cycling, we quantified the amount of product using denaturing PAGE with a constant voltage of \SI{200}{V} in a chamber immersed in a water bath at \SI{70}{\degreeCelsius}.
The gel contains 12.5\% 19:1 acrylamide/bis-acrylamide \SI{7}{M} urea, \SI{120}{mM} EDTA, and 1$\times$ TBE. 
The gel was stained with 1× SYBR Gold (Thermo Fisher Scientific) dissolved in 1$\times$ TBE buffer. Gel images were captured using a 16-bit fluorescence scanner (LI-COR)~[{\SFigPage}a~\cite{Note1}].

The gel images were processed and analyzed using a lab-developed Python code using the OpenCV library \cite{OpenCV}. Each lane was cropped into a rectangular region, and the fluorescence intensity of each pixel was averaged along the direction orthogonal to the lane. This yielded a one-dimensional intensity profile for each lane~[{\SFigPage}b~\cite{Note1}].
To remove background noise, we applied the rolling minimum method~\cite{yaroshchyk2014a} and subtracted the resulting background signal from the raw intensity data. The remaining signal peaks, corresponding to DNA products, were fitted using a Gaussian mixture model. The intensity of each peak was quantified as the area under the corresponding Gaussian curve.
A calibration curve relating fluorescence intensity to DNA mass was generated in each gel by using reference bands [{\SFigPage}~\cite{Note1}]. 

After imaging the gel, DNA was extracted from each product band. The gel around a target band was excised and transferred into microtubes (DNA LoBind, Eppendorf), crushed, mixed with ultrapure water, and incubated at \SI{37}{\degreeCelsius} for 1 hour in a thermostatic shaking incubator. The supernatant was collected for qPCR measurement~[Sec.~\ref{SSec:qPCR}].

\subsection{Measurement of error ratio}
\label{SSec:qPCR}
We quantified the error ratio by performing qPCR measurements on the supernatant obtained from the DNA extraction step using a real-time PCR cycler (CFX96, Bio-Rad).
We assumed that the concentration ratio between the correct and wrong products in the supernatant remained unchanged from the original sample prior to PAGE.
The qPCR reaction mixture contains the supernatant sample, \SI{200}{nM} of each primer~[{\STabPrimers}~\cite{Note1}], and Luna Universal qPCR Master Mix (New England Biolabs). The thermal cycling protocol consists of an initial denaturation step at \SI{95}{\degreeCelsius} for \SI{10}{s}, followed by an annealing step at either \SI{65}{\degreeCelsius} or \SI{50}{\degreeCelsius} for \SI{20}{s}, and an extension step at \SI{72}{\degreeCelsius} for \SI{5}{s}. 
The fluorescence data were analyzed to determine the threshold cycle (Ct values) by the method described in~\cite{boggy2010}.

For each experimental condition, we established independent calibration curves by measuring the Ct values of reference template strands with various concentrations prepared by serial dilutions with a ratio of 1/8~[{\SFigqPCR}; {\STabDNATemplates}~\cite{Note1}]. 
The Ct values were fitted by $\mathrm{Ct} = - a \log_8 (c/c_0) + b$, where $c$ is the template concentration, and $c_0=\SI{1}{pM}$ is the reference concentration. 
$a$ and $b$ are the fitting parameters~[{\STabDNACombinations}~\cite{Note1}].

\subsection{Melting curve experiment}
\label{SSec:Melting}
The melting curve experiment was conducted to determine the thermodynamic parameters of DNA hybridization for the simulations. 
We mixed DNA strands (\SI{100}{\nano M} for each strand), 0.25$\times$ SYBR Green I fluorescent dye for monitoring the concentrations of double-stranded DNA, and 1$\times$ Taq DNA Ligase Reaction Buffer (New England Biolabs), consistent with the conditions used in the main experiments. 
To prevent nonspecific hybridizations, the tag sequences were not used for the monomer strands~[{\STabMeltSec}~\cite{Note1}].
We gradually decreased the temperature of the sample in a stepwise manner from \SI{90}{\degreeCelsius} to \SI{30}{\degreeCelsius} by $\SI{0.2}{\degreeCelsius}$ every \SI{15}{s} and measured the fluorescence in each step [{\SFigMeltanAlysis}~\cite{Note1}].

The resulting fluorescence data were processed by using the exponential background removal method~\cite{palais2009}. 
The normalized melting curves $\alpha (T)$ were obtained by linearly scaling the background-corrected fluorescence data between 0 and 1 using the maximum and minimum fluorescence values. Normalization was carried out within a manually selected temperature range that encompassed the melting transition of each curve.
The standard free energy change, $\Delta G^\circ (T)$, was estimated from the normalized fluorescence signal $\alpha (T)$. Assuming that $\alpha (T)$ [{\SFigMeltanAlysis}b~\cite{Note1}] corresponds to the fraction of double-stranded DNA, $\Delta G^\circ (T)$ was calculated as
$\Delta G^\circ (T) = \kb T \ln \left[ 2c (1-\alpha (T))^2/\alpha (T) \right]$,
where $c=\SI{100}{nM}$ is the concentration of each DNA strand. $\Delta G^\circ(T)$ was then fitted by a function $\Delta H^\circ - T\Delta S^\circ$ to determine the standard enthalpy $\Delta H^\circ$ and the standard entropy $\Delta S^\circ$, which are assumed to be independent of $T$. This leads to the dissociation constant $K_\D = k_\off/k_\on = \exp \left( \Delta H^\circ/\kb T - \Delta S^\circ/\kb\right)$. The results are summarized in {\STabMeltRes}~\cite{Note1}.

We observed a linear dependence of $\Delta H^\circ$ and $\Delta S^\circ$ on the strand length $i$ for both correct and wrong substrates [{\SFigLengthDependence}~\cite{Note1}]; $\Delta H^\circ_\mathrm{R/W}(i) = \phi_\mathrm{H}\, i + \psi_\mathrm{H, R/W}$ and  
    $\Delta S^\circ_\mathrm{R/W}(i) = \phi_\mathrm{S}\,  i + \psi_\mathrm{S, R/W}$. The subscripts $\mathrm{R}$ and $\mathrm{W}$ refer to correct and wrong monomers.
The slopes ($\phi_\mathrm{H}$ and $\phi_\mathrm{S}$) were chosen in common between the correct and wrong monomers, representing the stabilization effect per sequence unit. 
The differences in intercepts ($\psi_\mathrm{H, W} - \psi_\mathrm{H, R}$ and $\psi_\mathrm{S, W} - \psi_\mathrm{S, R}$) represent the thermodynamic destabilization effect caused by hybridization error. 
The results are summarized in {\STabMeltRes}~\cite{Note1}.

\subsection{Numerical simulations}
\label{SSec:Simulation}

We constructed a model based on ordinary differential equations (ODEs) to simulate the experimental results, which incorporated hybridization, dissociation, ligation, and thermal cycling.

We modeled hybridization as a second-order reaction with an association rate constant $k_\on$, which was assumed to be independent of DNA length \cite{tawa2005} and temperature \cite{zhang2018a}. Up to three monomers can bind to the template strand.

The dissociation rate constant is denoted as $k_{\off,i,j}$ where $i$ represents the hybridization length in the unit of substrate monomers ($1 \leq i \leq l$), and $j$ ($0 \leq j \leq i$) denotes the number of errors in the substrate sequence. $l$ is the template length.
  The dissociation rate is given by: 
\begin{align}
    k_{\off,i,j}=k_\on K_{\D,i,j} = k_\on \exp \left(\Delta G_{i,j}^\circ/\kb \Tanneal \right),
\end{align}
where $\Delta G_{i,j}^\circ$ is given by
\begin{align}
    \Delta G_{i,j}^\circ = \Delta G_{\mono,\R}^\circ +(i-1)\left( \phi_\mathrm{H} + \Tanneal \phi_\mathrm{S} \right)  + j \Delta \Delta G.
\end{align} 
$\Delta \Delta G=\Delta G_{\mono, \W}^\circ - \Delta G_{\mono, \R}^\circ$ originates from the destabilization by a single error. 
We denote $\Delta G_{1, 0}^\circ$ and $\Delta G_{1,1}^\circ$ as $\Delta G_{\mono, \R}^\circ$ and $\Delta G_{\mono, \W}^\circ$, respectively.
The values of $\Delta G^\circ_\mathrm{mono,R/W}=\Delta H^\circ_\mathrm{R/W}(i=1)- \Tanneal\Delta S^\circ_\mathrm{R/W}(i=1)$, $\phi_\mathrm{H}$, and $\phi_\mathrm{S}$ were obtained from the melting curve experiments. 

The rate constants of the ligation $k_{\lig}$ were chosen depending on the sequences of the terminals to be ligated. For example, $k_{\lig,\dna{AB}}$ is used for the ligation of $\dna{A}$ and $\dna{BC}$, and $k_{\lig, \dna{BC}}$ is used for the ligation of $\dna{A'B}$ and C.
For simplicity, we assumed that $k_\lig$ is independent of temperature.

An example of the rate equation is 
\begin{align}
    &\frac{\D [\mathrm{A,B:\cdna{ABC}}]}{\D t} = k_\on [\dna{A}:\cdna{ABC}][\dna{B}] + k_\on [\dna{A}][\dna{B}:\cdna{ABC}] \nonumber \\ &+ k_{\off,1,0}[\dna{A,B,C}:\cdna{ABC}] - k_\on[\dna{A,B}:\cdna{ABC}][\dna{C}] \nonumber \\ &- 2k_{\off,1,0}[\dna{A,B}:\cdna{ABC}] - k_{\lig,\dna{AB}} [\dna{A,B}:\cdna{ABC}]. 
\end{align}
Here, colon (:) indicates the hybridized complex; e.g. $\dna{A,B}:\cdna{ABC}$ is the complex such that A and B are hybridized on a template $\cdna{ABC}$.

We simulated the thermal cycling by repeating the calculation of the dynamics for a duration of $\tau=20$ \si{s} at $\Tanneal$ and the disssociation of all the double-stranded DNAs, which corresponds to the denaturing.

\subsection{Fitting of the dynamics data in experiments by simulation curves}
\label{SSec:FittingProcedure}
To test the validity of our simulation, we fitted the simulation curves to the experimental curves of the dynamics measurement [Fig.~\ref{Fig:2}a].
The fitting parameters are $k_\on$, $\Delta G^\circ_\mathrm{mono, W}$, $k_{\lig,\dna{AB}}$, $k_{\lig,\dna{AB'}}'$, $k_{\lig,\dna{BC}}$, and $k_{\lig,\dna{B'C}}'$, which will be collectively denoted as $p$. 
We obtained $\Delta G^\circ_\mathrm{mono, W}$ as a fitting parameter, since the evaluation of $\Delta G^\circ_\mathrm{mono, W}$ from the melting curve is not simple, unlike $\Delta G^\circ_\mathrm{mono, R}$, due to the significant difference between $\Tanneal=\SI{66.0}{\degreeCelsius}$ and the melting temperature of $\dna{B'}:\cdna{B}$ ($\Tm=\SI{51.8}{\degreeCelsius}$).
The temperature dependence of $\Delta H^\circ$ is negilible only when $\Tanneal$ and $\Tm$ are similar when evaluating $\Delta G^\circ$.

We numerically integrated the rate equations and fitted the simulation curves to the experimental results by minimizing a cost function defined by
\begin{align}\label{eq:cost function}
    & C(\{\mu \},\{\sigma \}|p) = \nonumber \\ &\sum\limits_{m} \left[ \left( \frac{\mu_{m, \mathrm{a}}-f_{m, \mathrm{a}}(p)}{\sigma_{m, \mathrm{a}}} \right)^2 \left( \frac{\mu_{m, \mathrm{e}}-f_{m, \mathrm{e}}(p)}{\sigma_{m, \mathrm{e}}} \right)^2 \right].
\end{align}
Here, $\mu_{m,\mathrm{a/e}}$ and $\sigma_{m,\mathrm{a/e}}$ are the averages and standard errors of the total concentration (a) and the error ratio of the product with length $m$ in experiments (b), respectively. $f_{m,\mathrm{a/e}}$ are the values of the numerical simulations.  The minimization was done using the Nelder-Mead method~\cite{Nelder1965} implemented by the Optimization.jl package of Julia \cite{vaibhav_kumar_dixit_2023_7738525}.

We obtained $k_\on=9.41 \times 10^6$ \si{M^{-1}s^{-1}}, $\Delta G^\circ_\mathrm{mono, W} = -8.15$ \si{kcal \, {mol}^{-1}}, $k_{\lig,\dna{AB}}=0.222$ \si{s^{-1}}, $k_{\lig,\dna{AB}'}'=0.0190$ \si{s^{-1}}, $k_{\lig,\dna{BC}}=0.207$ \si{s^{-1}}, and $k_{\lig,\dna{BC}'}'=0.00184$ \si{s^{-1}}. We discuss the validity of the obtained parameters below.
The value of $k_\on$ was in the order of $10^6$, which aligns with the experimentally obtained values reported in some previous literature~\cite{zhang2018a, weitz2014a, yurke2003, aoyanagi2023}.
It is difficult to validate the ligation rates based on reference values because the actual concentration of active ligase is not accessible in experiments, and the ligation rate of the Taq DNA ligase with internal mismatches has not been reported for the present experimental condition. 
On the other hand, the specificity ratios, defined by the ratio of the correct and wrong ligation rates, were $k_{\lig,\dna{AB}}/k_{\lig,\dna{AB'}}' = 11.7$, and $k_{\lig,\dna{BC}}/k_{\lig,\dna{B'C}}' = 113$.
These values are compatible with the previous report that the specificity ratios are $\ge 8$ for an internal mismatch at various mismatch positions~\cite{pritchard1997}. 
The fact that the value for $\dna{B'+C}$ is approximately ten times larger than that for $\dna{A+B'}$ could be attributed to the nature of ligase that the mismatch at the 3' side has a larger discrimination ability than the mismatch at the 5' side, as was previously reported~\cite{luo1996a}.

\subsection{Simulations of individual-step experiments}
\label{SSec:TempDepMid}
We simulated the experiments in Fig.~\ref{Fig:3} by using the same model as in the dynamics experiment [Sec.~\ref{SSec:FittingProcedure}].
For the middle errors [Fig.~\ref{Fig:3}a], we used the parameters described in Sec.~\ref{SSec:FittingProcedure}.
For the terminal errors [Fig.~\ref{Fig:3}b],  we need additional parameter $k_{\lig,\dna{A'B}}'$, which was estimated to be $k_{\lig,\dna{A'B}}'= 5.51\times10^{-4}$ by manual fitting of the experimental results in Fig.~\ref{Fig:3}b by simulations.
This value is significantly smaller than $k_{\lig,\dna{AB'}}'$ and $k_{\lig,\dna{B'C}}'$. 
This difference could be caused by the geometry of the error site.
As noted above, the specificity ratio increases when the error is positioned at the 3' side of the ligation site~\cite{luo1996a}.

\subsection{Simulations with long templates}
\label{SSec:SingleSite}

The template concentration was fixed at \SI{1}{nM} and the total substrate concentration at each binding domain was set to \SI{40}{nM}. To maintain this total, we mixed \SI{20}{nM} of the correct monomer with \SI{20}{nM} of the wrong monomer. The hybridization rate $k_\on$ was set to $10^7\, \si{/M/s}$. 
For simplicity, $k_\lig$ was fixed at \SI{10}{s^{-1}} across all cases, regardless of whether an error was present, its position, or the length of the substrate. 
We stopped the simulations when the total concentration of final products, including both correct and wrong sequences, reached a specific threshold value, which was determined depending on $\Delta G_{\mono, \R}^\circ$; 0.0001 \si{pM} for $\Delta G_{\mono, \R}^\circ = - 5\,\kb T$ and 1 \si{pM} for $- 10\,\kb T$.
The lengh-dependent error ratio was calculated as $\varepsilon (l)= w_l/ r_l $ where $r_l$ and $w_l$ are the concentrations of the correct and wrong final products, respectively, with the length of $l$.

For the chain-growth polymerization, we set $k_{\off,i,j} = 0$ for $i \geq 2$, allowing only monomer hybridization and dissociation while preventing intermediate dissociation.

For the simulations in Fig.~\ref{Fig:5}e, we used the same setup as in the single-site error simulations. The only difference is that the errors are allowed at any site.

\begin{acknowledgments}
We thank Yohei Nakayama for the helpful discussion.
This work is supported by JSPS KAKENHI Grant Number JP23K17658 and JST ERATO Grant Number JPMJER2302 (all to S.T.), and JST SPRING Grant Number JPMJSP2114, Japan (to H.A.).
\end{acknowledgments}

%


\clearpage

\begin{widetext}

\textbf{\Large Supplementary Materials}\\

\renewcommand{\thefigure}{S\arabic{figure}}
\renewcommand{\thesection}{S\arabic{section}}
\renewcommand{\thetable}{S\arabic{table}}
\renewcommand{\theequation}{S\arabic{equation}}
\renewcommand{\thesubsection}{S\arabic{section}.\arabic{subsection}}

\setcounter{section}{0}  
\setcounter{subsection}{0} 
\setcounter{subsubsection}{0} 
\setcounter{figure}{0}  
\setcounter{table}{0}  

\section{Supplementary Tables}

\begin{table}[htbp]
\centering
\caption{DNA sequences used in the ligation experiment. [PHO] denotes a phosphate group.
U, $\dna{U'}$, X, and Y are tag sequences for qPCR probing.}
\begin{tabular}{l|l}
\hline
Name & Sequence $(5' \rightarrow 3')$\\
\hline
\hline
$\cdna{ABC}$ & 
\begin{tabular}{l} gtgaatt aaccaggaagtgagacgaa \\ taacgtaggtgttggctcta  tatccatcacgactcagcaaa  tgtactg \end{tabular} 
\\ 
\hline
U-A & 
aata gcataccgatgcttgaccacat  attgctgagtcgtgatggat \\ 
\hline 
$\dna{U'}$-$\dna{A'}$ & 
aata gcataccgatgcttgaccacgc  attgctgggtcgcgatggat \\
\hline
B & 
[PHO]-atagagccaacacctacgtt \\ 
\hline
$\dna{B'}$ & 
[PHO]-atagagctaacatctacgtt \\ 
\hline
C-X & 
[PHO]-attcgtctcacttcctggtt tcggacagtctgctacagcg \\ 
\hline 
BC-X & [PHO]-atagagccaacacctacgtt attcgtctcacttcctggtt tcggacagtctgctacagcg \\ 
\hline
$\dna{B'C}$-Y & [PHO]-atagagctaacatctacgtt attcgtctcacttcctggtt tacacatgacgcacggatgc \\ 
\hline
U-AB & gcataccgatgcttgaccacat attgctgagtcgtgatggat atagagccaacacctacgtt \\
\hline
$\dna{U'}$-$\dna{A'B}$ & gcataccgatgcttgaccacgc attgctgggtcgcgatggat atagagccaacacctacgtt \\
\hline
\end{tabular}
\label{tab:seqlpcr}
\end{table}

\begin{table}[htbp]
\centering
\caption{Primer sequences used in qPCR.}
\begin{tabular}{l|l}
    \hline
    Name & Sequence $(5' \rightarrow 3')$ \\
    \hline
    \hline
    p-U &  gcataccgatgcttgaccacat \\
    \hline
    p-$\dna{U'}$ & gcataccgatgcttgaccacgc \\
    \hline
    p-X & cgctgtagcagactgtccga \\ 
    \hline
    p-Y & gcatccgtgcgtcatgtgta \\
    \hline
    p-ABC & ggaagtg agacg aattt cgtagg tgttg \\
    \hline 
    p-$\dna{AB'C}$ & ggaagtg agacg aattt cgtaga tgtta \\
    \hline
    p-BCX & atggatttagagccaacac \\
    \hline
    p-$\dna{B'CX}$ & tgagtcgtgatggat taagagctaacat \\
    \hline
    p-B & atagagccaacacctacgt \\ 
    \hline
\end{tabular}
    \label{tab:primers}
\end{table}

\begin{table}[htbp]
\centering
\caption{DNA sequences of the templates used for calibration curve measurement.}
\begin{tabular}{l|l}
    \hline
    Name & Sequence $(5' \rightarrow 3')$ \\
    \hline 
    \hline
    t-U-X &  gcataccgatgcttgaccacat  attgctgagt  cttcctggtt tcggacagtctgctacagcg \\
    \hline
    t-trimer($\dna{B'}$) & gcataccgatgcttgaccacat  attgctgagtcgtgatggat atagagctaacatctacgtt attcgtctcacttcctggtt \\
    \hline
    t-trimer($\dna{A'}$) & gcataccgatgcttgaccacgc  attgctgggt  cttcctggtt tcggacagtctgctacagcg \\
    \hline
    t-U-Y & gcataccgatgcttgaccacat  attgctgagt  cttcctggtt tacacatgacgcacggatgc \\ 
    \hline
    t-UAB & gcataccgatgcttgaccacat  attgctgagtcgtgatggat atagagccaacacctacgtt \\
    \hline
    t-$\dna{UAB'}$ & gcataccgatgcttgaccacat  attgctgagtcgtgatggat atagagctaacatctacgtt \\
    \hline 
    t-BCX & atagagccaacacctacgttattcgtctcacttcctggtttcggacagtctgctacagcg \\
    \hline
    t-$\dna{B'CX}$ & atagagctaacatctacgttattcgtctcacttcctggtttcggacagtctgctacagcg \\
    \hline
    t-$\dna{U'A'B}$ & gcataccgatgcttgaccacgc  attgctgggtcgcgatggat atagagccaacacctacgtt \\
    \hline
\end{tabular}
    \label{tab:templates}
\end{table}

\begin{table}[htbp]
\centering
\caption{Combinations of primer sets and templates used for calibration curve measurements, and the corresponding calibration parameters $a$ and $b$ (see Sec.~V D).}
\begin{tabular}{l|cccccc}
    Products & Forward primer & Reverse primer & Template & Annealing temp. (\si{\degreeCelsius}) & $a$ & $b$\\
    \hline 
    \hline
    UABCX+$\dna{UAB'CX}$ &  p-U & p-X & t-U-X & 65 & -2.87 & 15.9 \\
    \hline
    $\dna{UAB'CX}$ & p-U & p-$\dna{AB'C}$ & t-trimer($\dna{B'}$) & 65 & -3.06 & 18.5\\
    \hline
    $\dna{U'A'BCX}$ & p-$\dna{U'}$ & p-X & t-trimer($\dna{A'}$) & 65 & -2.93 & 17.2 \\
    \hline
    UABCY & p-U & p-Y & t-U-Y & 65 & -3.00 & 17.4 \\
    \hline
    UAB & p-U & p-ABC & t-UAB & 50 & -2.92 & 19.2 \\ 
    \hline
    $\dna{UAB'}$ & p-U & p-$\dna{AB'C}$ & t-$\dna{UAB'}$ & 50 & -3.08 & 24.5 \\
    \hline
    UAB & p-U & p-B & t-UAB & 65 & -2.88 & 16.9 \\  
    \hline
    $\dna{U'A'B}$ & p-$\dna{U'}$ & p-B & t-$\dna{U'A'B}$ & 65 & -2.89 & 16.9 \\
    \hline
    BCX & p-BCX & p-X & t-BCX & 50 & -3.09 & 17.1 \\
    \hline 
    $\dna{B'CX}$ & p-$\dna{B'CX}$ & p-X & t-$\dna{B'CX}$ & 50 & -3.31 & 22.4 \\
    \hline
\end{tabular}
    \label{tab:combinations}
\end{table}

\begin{table}[htpb]
    \centering
    \caption{DNA sequences used in the melting curve experiment.}
    \begin{tabular}{l|l}
        Name & Sequence $(5' \rightarrow 3')$ \\
        \hline
        \hline
        $\cdna{A}$ & tatccatcacgactcagcaa \\
         \hline
         $\cdna{B}$ & taacgtaggtgttggctcta \\
         \hline
         $\cdna{C}$ & taaccaggaagtgagacgaat  \\
         \hline
         $\cdna{AB}$ & taacgtaggtgttggctcta tatccatcacgactcagcaa \\
         \hline
         $\cdna{BC}$ & taaccaggaagtgagacgaa taacgtaggtgttggctctat \\
         \hline
         A & attgctgagtcgtgatggat \\
         \hline
         B & [PHO]-atagagccaacacctacgtt \\
         \hline
         $\dna{B'}$ & [PHO]-atagagctaacatctacgtt \\
         \hline
         C & attcgtctcacttcctggtt \\
         \hline
         UAB & gcataccgatgcttgaccacat  attgctgagtcgtgatggat atagagccaacacctacgtt \\
         \hline
         $\dna{UAB'}$ & gcataccgatgcttgaccacat  attgctgagtcgtgatggat atagagctaacatctacgtt \\
         \hline
         BCX & atagagccaacacctacgtt attcgtctcacttcctggtt tcggacagtctgctacagcg \\
         \hline
         $\dna{B'CX}$ & atagagctaacatctacgtt attcgtctcacttcctggtt tcggacagtctgctacagcg \\
         \hline
    \end{tabular}
    \label{tab:sequencemelt}
\end{table}

\begin{table}[htbp]
\centering
\caption{Thermodynamic parameters obtained from the melting curve analysis. The average was taken over three independent experiments for each pair.
}
\begin{tabular}{ccccc}
Pair & $\Tm$ (\si{\degreeCelsius}) & $\Delta H^\mathrm{o}$ (\si{k cal/mol}) & $\Delta S^\mathrm{o}$ (\si{cal/mol}) & $K_\D$ at \SI{66.0}{\degreeCelsius} (nM) \\
\hline
\hline
$\dna{A}:\cdna{A}$ & 64.9 & -177 & -493 & 483 \\
\hline
$\dna{B}:\cdna{B}$ & 64.5 & -180 & -502 & 663\\
\hline
$\dna{B'}:\cdna{B}$ & 51.8 & -149 & -427 & 3040000 \\
\hline
$\dna{C}:\cdna{C}$ & 64.1 & -190 & -532 & 977 \\
\hline
$\dna{AB}:\cdna{AB}$ & 75.8 & -319 & -883 & 0.000358 \\
\hline 
$\dna{AB'}:\cdna{AB}$ & 70.6 & -279 & -781 & 0.752 \\
\hline
$\dna{BC}:\cdna{BC}$ & 75.6 & -337 & -936 & 0.000209 \\
\hline 
$\dna{B'C}:\cdna{BC}$ & 70.2 & -299 & -840 & 0.828  \\
\hline
Correct monomer & 64.5 & -183 $(\Delta H_\mathrm{mono,R}^\circ)$ & -511 $(\Delta S_\mathrm{mono,R}^\circ)$ & 640 \\
\hline
Wrong monomer & 52.4 & -147 $(\Delta H_\mathrm{mono,W}^\circ)$ & -419 $(\Delta S_\mathrm{mono,W}^\circ)$ & 1790000 \\
\hline
\end{tabular}
\label{tab:tmkd}
\end{table}

\clearpage
\section{Supplementary Figures}

\begin{figure}[h!]
     \centering
     \includegraphics{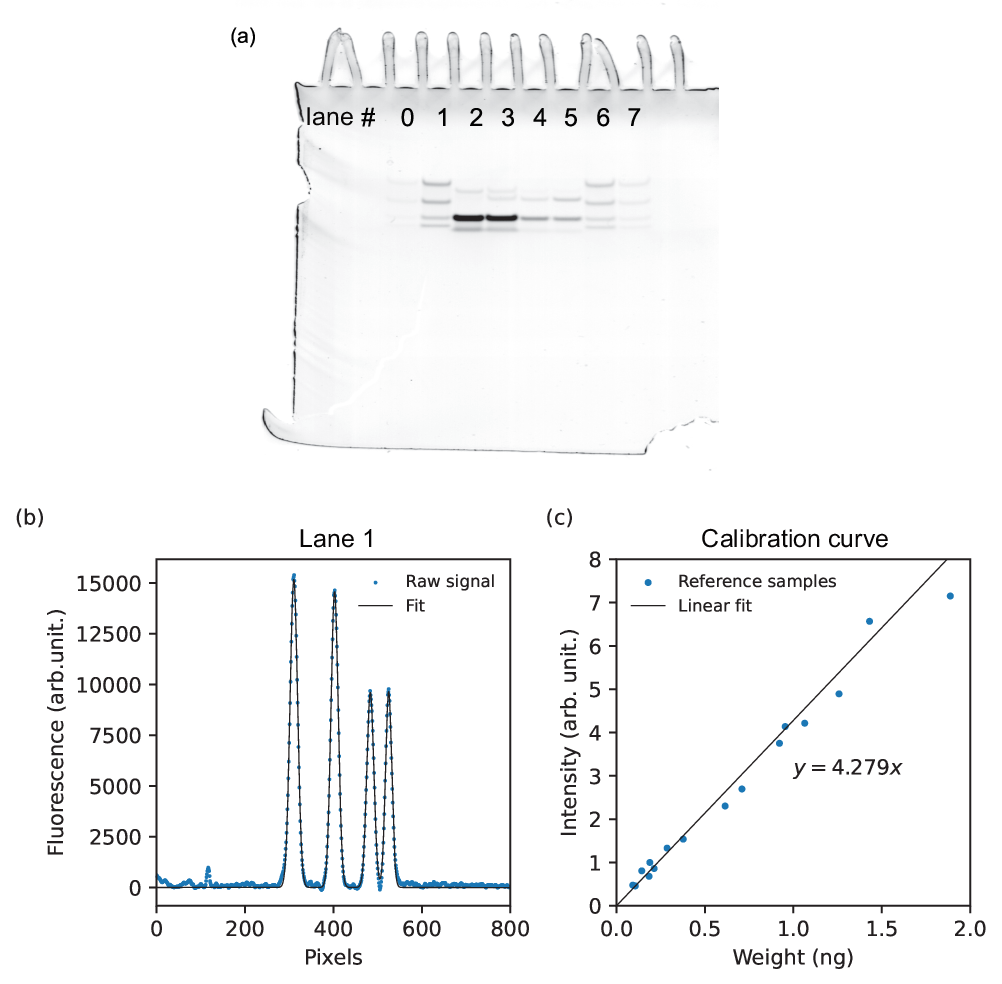}
     \caption{Quantification by denaturing PAGE. (a) An example of the gel image. A reference DNA mixture containing single-stranded DNAs of 82, 60, 46, and 40 nucleotides in length was loaded into the outermost lanes (0, 1, 6, and 7) at both ends of the gel with different amounts (15, 150, 100, and 30 fmol for each band in lanes 0, 1, 6, and 7, respectively). (b) Intensity profile of a reference lane (lane 1) of the gel in (a) after subtracting the background signal (symbols) and the fitting curve by a Gaussian mixture (solid curve). (c) Calibration curve. For each reference lane, we extracted the 1D fluorescence profile as in (b) and calculated the area under each Gaussian peak. Fluorescence intensities were then associated with the known DNA weights. Dots represent reference data points, and the straight line indicates the linear fit with a fixed intercept at zero.}
     \label{fig:page}
 \end{figure}

 \begin{figure}[h!]
    \centering
    \includegraphics{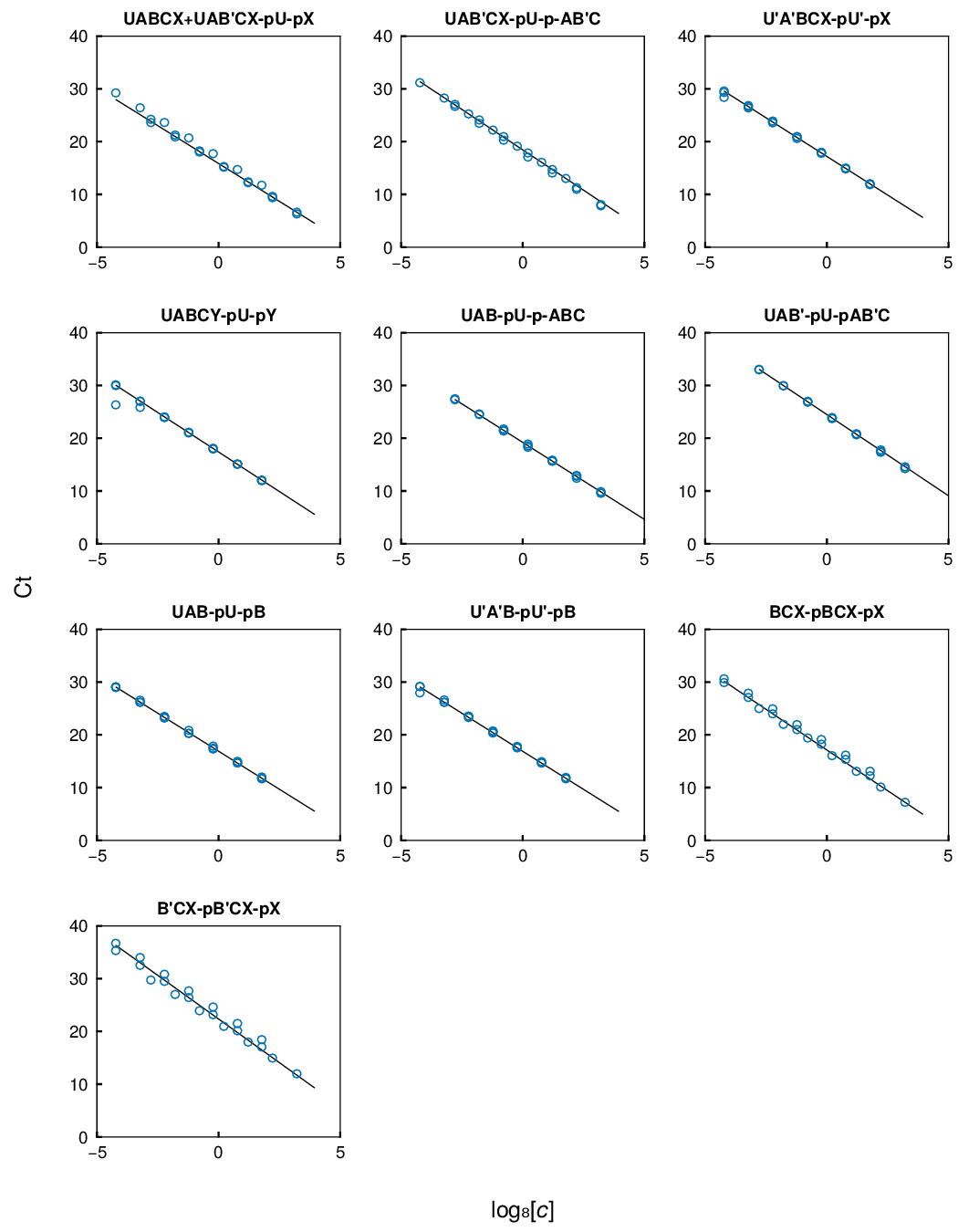}
    \caption{Calibration curves used for quantifying the error ratio. For each condition, measurements were repeated three times, and the data were jointly fitted using the method described in Sec.~V~D. The parameters obtained are listed in Table~\ref{tab:combinations}.}
    \label{fig:calibrationcurves}
\end{figure}

\begin{figure}[h!]
    \centering
    \includegraphics{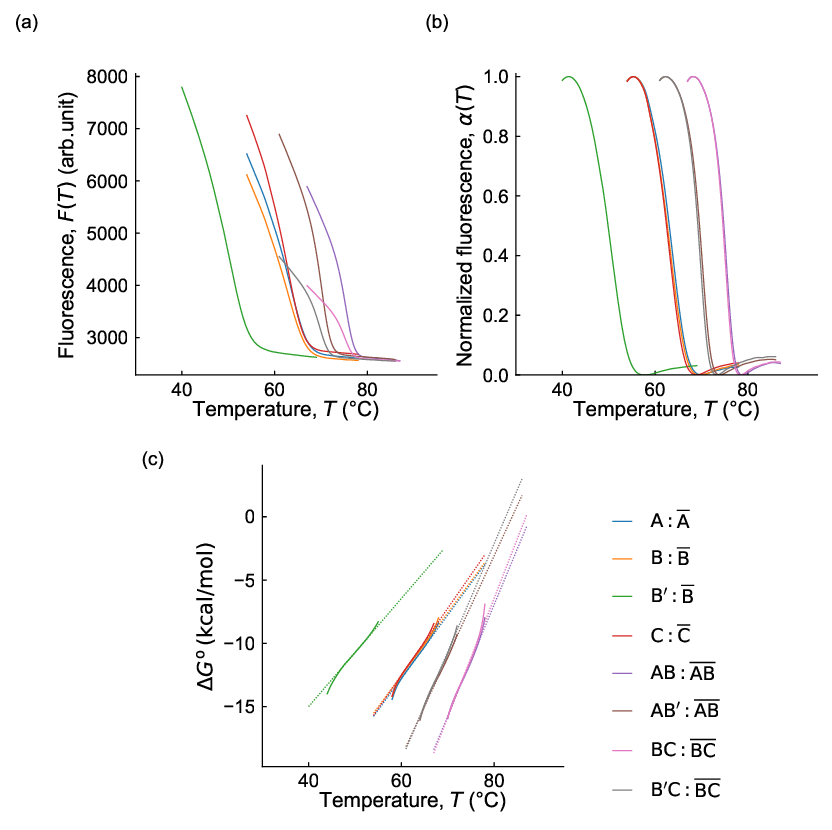}
    \caption{(a) The raw fluorescence data $F(T)$. (b) The normalized melting curves $\alpha(T)$ was obtained by substracting an exponential background from $F(T)$ \cite{palais2009} and then normalizing it by scaling the value so that the maximum and minimum become 1 and 0, respectively, within a temperature range individually chosen for each curve. (c) Standard free energy change associated with hybridization calculated as $\Delta G^\circ (T)=\kb T \ln \left[ 2c(1-\alpha(T) )^2/\alpha(T) \right]$ with $c$ denoting the concentration of each DNA strand. The dashed lines in (c) are linear fits by a function $\Delta H^\circ - T\Delta S^\circ$. 
    The fitting was done in the range of $ \SI{\pm 3}{\degreeCelsius}$ around the temperature where $\alpha(T)=0.5$.    
    Estimated parameters $\Delta H^\circ$ and $\Delta S^\circ$ are listed in Table \ref{tab:tmkd}.}
    \label{fig:meltanalysis}
\end{figure}

\begin{figure}[h!]
    \centering
    \includegraphics{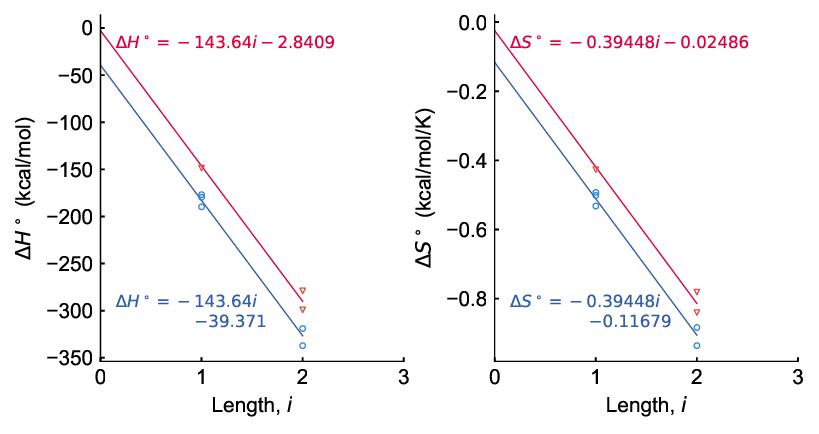}
    \caption{Length dependence of $\Delta H^\circ$ and $\Delta S^\circ$ for the hybridizations with correct pairs (blue circle) and wrong pairs (red triangle) and their linear fits (solid curves). 
    A common slope is used for both fitting conditions.}
    \label{fig:lengthdependence}
\end{figure}

\begin{figure}[h!]
    \centering
    \includegraphics{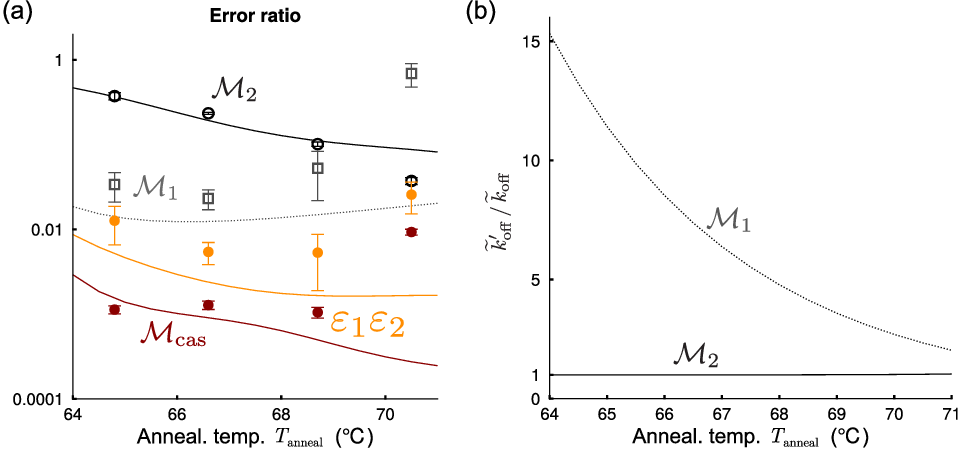}
    \caption{(a) Temperature dependence of the error ratio when a wrong substrate $\dna{B'}$ is incorporated in the middle of the template. Error bars represent standard errors. Yellow symbols and solid curves indicate the product $\error_1 \error_2$, which closely approximates the observed error ratio in cascade replication, $\error_\mathrm{cas}$. Error bars for $\error_1 \error_2$ were calculated by propagating the errors of $\error_1$ and $\error_2$. (b) The ratio $\widetilde k_\off'/\widetilde k_\off$ is calculated for various temperatures.
    }
    \label{fig:multiply_center}
\end{figure}

\begin{figure}[h!]
    \centering
    \includegraphics{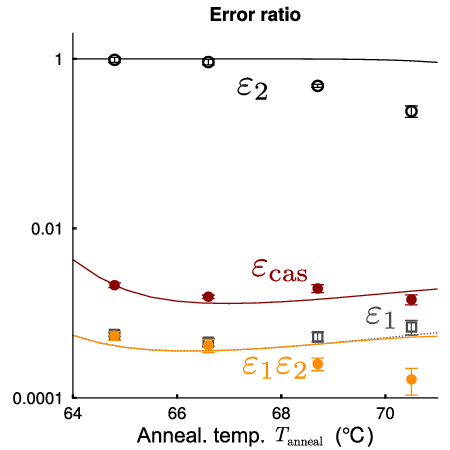}
    \caption{Temperature dependence of the error ratio when the error is introduced at the terminal position of the template. Error bars represent standard error. The yellow points and line denote the expected error ratio based on the multiplication $\error_1 \error_2$, which approximates the measured cascade error ratio, $\error_\mathrm{cas}$. Same as Fig.~\ref{fig:multiply_center}, error bars in $\error_1 \error_2$ was estimated by propagating the standard errors of $\error_1$ and $\error_2$.}
    \label{fig:multiply_edge}
\end{figure}

\end{widetext}

\end{document}